\documentclass[10pt,aps,pre,twocolumn,nofootinbib,superscriptaddress,longbibliography]{revtex4-1}
\usepackage[colorinlistoftodos,textsize=tiny]{todonotes}
\usepackage[english]{babel}
\usepackage[utf8x]{inputenc}
\usepackage{amsmath,amsfonts,amssymb,amsthm}
\usepackage{graphicx}
\usepackage{hyperref}
\usepackage{color}
\usepackage{hyphenat}
\usepackage{natbib}
\usepackage{microtype}
\usepackage{mathpazo}
\usepackage{cleveref}
\usepackage[labelformat = empty,position=top]{subcaption}
\usepackage[export]{adjustbox}
\usepackage{pgfplots}
\usepackage{pgfgantt}
\usepackage{mathtools}
\usepackage[normalem]{ulem} 
\usepackage{tikz}
\usetikzlibrary{calc}
\pgfplotsset{compat=newest}
\pgfplotsset{plot coordinates/math parser=false}
\makeatletter
\renewcommand*\env@matrix[1][\arraystretch]{%
  \edef\arraystretch{#1}%
  \hskip -\arraycolsep
  \let\@ifnextchar\new@ifnextchar
  \array{*\c@MaxMatrixCols c}}
\makeatother

\newcommand{\brho}{\boldsymbol \rho}
\newcommand{\bsigma}{\boldsymbol \sigma}
\newcommand{\btheta}{\boldsymbol \theta}

\newcommand{\bI}{\mathbf{I}}
\newcommand{\bL}{\mathbf{L}}
\newcommand{\bA}{\mathbf{A}}
\newcommand{\bD}{\mathbf{D}}
\newcommand{\Tr}[1]{\operatorname{Tr}\left\lbrack #1\right\rbrack}
\newcommand{\pin}{p_{\textrm{in}}}
\newcommand{\pout}{p_{\textrm{out}}}

\newcommand{\argmin}{\operatorname{argmin}}
\newcommand{\bra}[1]{\left \langle #1 \right|}
\newcommand{\ket}[1]{\left | #1 \right \rangle}

\Crefname{equation}{Eq.}{Eqs.}

\pgfdeclarelayer{background}
\pgfdeclarelayer{foreground}
\pgfsetlayers{background,main,foreground}

\tikzset{declare function={polyradius(\numnodes)=1/(2*sin(pi/\numnodes));}}

\begin{document}
\title{Thermodynamics of network model fitting with spectral entropies}
\author{Carlo Nicolini}
\affiliation{Center for Neuroscience and Cognitive Systems, Istituto Italiano di Tecnologia, Corso Bettini 31, 38068 Rovereto (TN), Italy}
\affiliation{These authors contributed equally to this work}
\author{Vladimir Vlasov}
\affiliation{Center for Neuroscience and Cognitive Systems, Istituto Italiano di Tecnologia, Corso Bettini 31, 38068 Rovereto (TN), Italy}
\affiliation{These authors contributed equally to this work}
\author{Angelo Bifone}
\affiliation{Center for Neuroscience and Cognitive Systems, Istituto Italiano di Tecnologia, Corso Bettini 31, 38068 Rovereto (TN), Italy}
\date{\today}
\begin{abstract}
An information theoretic approach inspired by quantum statistical mechanics was recently proposed as a means to optimize network models and to assess their likelihood against synthetic and real-world networks.
Importantly, this method does not rely on specific topological features or network descriptors, but leverages entropy-based measures of network distance.
Entertaining the analogy with thermodynamics, we provide a physical interpretation of model hyperparameters and propose analytical procedures for their estimate.
These results enable the practical application of this novel and powerful framework to network model inference.
We demonstrate this method in synthetic networks endowed with a modular structure, and in real-world brain connectivity networks.

\end{abstract}
\maketitle

\section{Introduction}
Many natural and artificial phenomena can be represented as networks of interacting elements.
The mathematical framework of network theory can be applied across disciplines, ranging from sociology to neuroscience, and provides a powerful means to investigate a variety of diverse phenomena~\cite{newman2010book}.
Unveiling the structural and organizational principles of complex networks often implies comparison with statistical network models.
Generative models, for example, describe mechanisms of network wiring and evolution~\cite{barabasi1999,caldarelli2007}, or the constraints that may have contributed to shaping the network topology during its development~\cite{bullmore2009}.
Null models are used to describe maximally random networks with specific features, for example prescribed sequences of node degrees~\cite{wasserman1994}.

Maximum likelihood approaches have been proposed to compare the ability of different models to describe real-world networks and to optimize model parameters to best fit experimental networks.
{These methods are designed to assign the same probability to networks satisfying the same set of constraint, but hardly can take into account the whole network structure~\cite{squartini2017book}.}

Recently, an information theoretic framework inspired by quantum statistical mechanics principles has been proposed as a tool to assess and optimize network models~\cite{dedomenico2016b}.
This approach relies on the minimization of the relative entropy based on the network spectral properties. Importantly, this relative entropy does not depend on a distribution of specific descriptors, but considers the network as a whole.
However, this representation introduces an external, tunable hyperparameter $\beta$: the optimal estimate from the relative entropy minimization procedure critically depends on the choice of $\beta$, a major limitation to the practical use of this framework. 

Relative entropy is a central concept in thermodynamics of information (see review~\cite{parrondo2015}) and is defined on the basis of the density matrix.
In light of this thermodynamic analogy, here we build a physical interpretation of this approach to network optimization and fitting, where $\beta$ plays the role of an inverse temperature.
This provides criteria for a rigorous selection of the optimal $\beta$, and
enables the practical application of relative entropy minimization in the optimal reconstruction of parameters for different network models.

The paper is structured as follows.
First, we present the theoretical framework of classical maximum entropy null models as a way to generate maximally random ensemble of networks with given constraints.
We move on to briefly discussing how the Erd\H{o}s-R{\'{e}}nyi random graph and the configuration model emerge naturally from these ideas.

Within the settings of spectral entropies, we propose a new practical optimization method based on an approximation of the Laplacian spectrum and give a concise closed form expression for the gradients of relative entropy with respect to the model parameters.
We continue discussing a thermodynamic interpretation of the meaning of relative entropy optimization in terms of irreversible processes.

We calculate analytically the optimal temperature parameter of the Erd\H{o}s-R{\'{e}}nyi and planted partition models.
Furthermore, we generalize this result to more complex models with the help of numerical simulations showing the advantages of the spectral entropy approach with respect to other maximum likelihood methods.

Finally, we demonstrate the use of spectral entropies for the optimization of a generative model of neural connectivity in a real-world dataset.

\section{Models of complex networks}\label{sec:modelsofcomplexnetworks}
We summarize here a few definitions which are necessary to make this paper self-contained.
Let us consider here simple binary undirected graphs $G=(V,E)$ with $|V|=n$ number of nodes and $|E|=m$ number of links.

The adjacency matrix is denoted as $\bA=\{a_{ij}\}$ and the (combinatorial) graph Laplacian as $\bL = \bD - \bA$, where $\bD$ is the diagonal matrix of the node degrees.
Notably, the (combinatorial) Laplacian matrix associated with an undirected graph is a semi-positive definite matrix, meaning that all its eigenvalues $\lambda_1 \geq \ldots \geq \lambda_n = 0$ are positive (or zero) and real.
A random graph model is an ensemble of networks randomly defined in the probability space $\Omega$ and distributed around some specific network property.
For example, in the Gilbert random graph model the probability distribution $\mathcal{P}(G)$ is sharply peaked in $\mathcal{P}(G)=1/\Omega$ for graphs with $n$ nodes and exactly $m$ edges, and is zero otherwise~\cite{newman2010book}.
This distribution is well described in statistical mechanics as the microcanonical ensemble, as it enforces the constraints strictly.

However, given the combinatorial complexity of dealing with the micro-canonical description, it is easier to fix the average value of observables of interest rather than working with exact constraints.
This approach gives rise to the canonical ensemble of random graphs~\cite{park2004,newman2010book}.
This type of models has the same role in the study of complex networks as the Boltzmann distribution in classical statistical mechanics; it gives the maximally uninformed prediction of some network properties subject to the imposed constraints.
These ideas can be dated back to the Jaynes' maximum entropy principle~\cite{jaynes1957b}.
In this sense, the maximally random ensemble of graphs satisfying the imposed topological constraints on average also takes the name of the Exponential Random Graph Model (ERGM)~\cite{newman2010book} or $p^*$-model in the social sciences~\cite{wasserman1994}.
In its simplest implementation, the ERGM results in the Erd\H{o}s-R{\'{e}}nyi~\cite{erdos1959} model, where the link probability is constant.

On the other hand, if one wants to generate the maximally random network that maintains the desired degree sequence~$\{ k_i\}$, the resulting ensemble is called the Undirected Binary Configuration Model (UBCM)~\cite{caldarelli2002,park2004,squartini2015}.
Being the degree an entirely local topological property, it is affected by the intrinsic properties of vertices.
For this reason, one can assign a hidden variable $x_i \geq 0$ to each node.
Its value acts as a fitness score, which is hypothesized to be proportional to the expected node degree~\cite{caldarelli2002}.
If two nodes have a high fitness score, they are more likely to be connected by a link.
In this model one can describe the link probability as the normalized product of their scores~\cite{park2004,squartini2014,squartini2015}, resulting in the following expected link probability:
\begin{equation}\label{eq:ubcm}
p_{ij} = {\mathbb{E}}[ a_{ij}] = \frac{x_i x_j}{1+x_i x_j}
\end{equation}
The values of $x_i$ are obtained by numerical optimization of a specifically designed likelihood function~\cite{squartini2015,garlaschelli2008}.
In this framework, the hidden variables $x_i$ are the Lagrange multipliers of the constrained problem that ensures the expected degree $\langle{k_i}\rangle=\sum_{i\neq j}\langle a_{ij}\rangle$ of the vertex $i$ equals on average its empirical value $k_i$.
Interestingly this model highlights the fermionic properties of the links, as they are modeled like particles with only two states, namely the link being present or not.

If the network is sufficiently random, the degree-sequence alone can model the higher order patterns like the clustering coefficient or the average nearest neighbor degree.
However, deviations of other graph theoretical measures between model and empirical network are indicative of genuine higher-order patterns, like clustering or rich clubs, not simply accountable by the degree sequence alone~\cite{squartini2014,garlaschelli2008}.

\section{Spectral entropies framework}
A measure of complexity is central to the understanding of differences and similarities between networks, and to decode the information that they represent.
Supported by the seminal demonstration that the von Neumann entropy of a properly defined density matrix may be used for network comparison~\cite{dedomenico2016b}, in this paper we address the unsolved problem of inverse temperature selection and show that the {result of} model fitting strongly depend on it.

The first observation that an appropriately normalized graph Laplacian can be treated as a density matrix of a quantum system, is credited to the authors of reference~\cite{braunstein2006a}.
Indeed, the Laplacian spectrum encloses a number of important topological properties of the graph~\cite{estrada2011,anderson1985,merris1994a,delange2014}.
For instance, the multiplicity of the zero eigenvalue corresponds to the number of connected components, the multiplicity of each eigenvalue is related to graph symmetries~\cite{merris1994a,delange2014,delange2016}, the concept of expanders and isoperimetric number are connected to the first and second largest eigenvalues~\cite{cheeger1970,donetti2006}.
Moreover, the graph Laplacian appears often in the study of random walkers~\cite{lovasz1993,masuda2017}, diffusion~\cite{bray1988}, combinatorics~\cite{mohar1991laplacian}{,} and a large number of other applications~\cite{merris1994a,mohar1991laplacian}.

After the first demonstration that a graph can be always represented as a uniform mixture of pure density matrices~\cite{braunstein2006a}, at least two different definitions of quantum density for complex networks have been used~\cite{anand2011,dedomenico2016b}.
Adopting the notation of quantum physics, the von Neumann density matrix $\brho$ is a Hermitian and positive definite matrix with unitary trace, that admits a spectral decomposition as:
\begin{equation}
\brho = \sum_{i=1}^n \lambda_i(\brho) \ket{\phi_i}\bra{\phi_i}
\end{equation}
for an orthonormal basis $\{\ket{\phi_i}\}$ and eigenvalues $\lambda_i(\brho)$. Thus, a density matrix can be represented as a convex combination of pure states~\cite{anand2011}.

The von Neumann entropy of the density operator $\brho$ can be expressed as the Shannon entropy of its eigenvalues~\cite{wilde2013}:
\begin{equation}\label{eq:vonneumannentropy}
S(\brho) = - \Tr{\brho \log \brho} = - \sum_{i=1}^n \lambda_i(\brho)  \log \lambda_i(\brho)
\end{equation}
where $\log(\boldsymbol \cdot)$ is the principal matrix logarithm~\cite{higham2008} when the argument is a matrix.
The von Neumann entropy of the density matrix is bounded between $0$ and $\log n$~\cite{wilde2013}.

We adopt the quantum statistical mechanics perspective~\cite{dedomenico2016b}, where the von Neumann density matrix $\brho$ of a complex network is built considering a quantum system with Hamiltonian $\bL$ in thermal contact with a heat bath at constant temperature $k_B T=1/\beta$, where $k_B$ is the Boltzmann constant.

{In the perspective of Jaynes' maximum entropy framework~\cite{jaynes2003}, the state of maximum uncertainty about the system, constrained by the conditions $\Tr{\brho}=1$ and $\langle \bL \rangle=\Tr{\brho \bL}$} is described by the quantum Gibbs-Boltzmann distribution:
\begin{equation}\label{eq:densitymatrix}
\brho = \frac{e^{-\beta \bL}}{\Tr{e^{-\beta \bL}}},
\end{equation}
where here $e^{(\boldsymbol \cdot)}$ is the matrix exponential when the argument is a matrix and the denominator is the so-called partition function of the system, i.e. the sum over all possible configurations of the ensemble, and is denoted by $Z=\Tr{e^{-\beta \bL}}$.
Borrowing the terminology of statistical physics, thermal averages of any graph-theoretical measure $\mathbf{O}$ over the ensemble defined by $\brho$ are obtained as: 
\begin{equation}\label{eq:traceorho}
\langle \mathbf{O} \rangle_{\brho} = \Tr{\mathbf{O} \brho} = \frac{1}{Z}\Tr{\mathbf{O} e^{-\beta \bL}}.
\end{equation}

The choice of this density matrix for complex networks is supported by the observation that previous definitions of entropy~\cite{braunstein2006a,anand2011} in graph theory resulted in violation of sub-additivity~\cite{dedomenico2016b,biamonte2017}, a central property of entropy~\cite{wilde2013}.
The strength of this definition of von Neumann entropy for graphs lies in the possibility to establish a connection between quantum statistical mechanics and the realm of networks.
Moreover, this approach closely resembles the one taken in the study of diffusion on networks~\cite{masuda2017}, where $\beta$ is no longer interpreted as an inverse temperature of the external heat bath, but rather as the diffusion time of a random walker~\cite{estrada2008,faccin2013}.
This renders the idea that the network properties can be explored at different scales by varying $\beta$~\cite{estrada2012a,biamonte2017,masuda2017}.

\section{Model optimization}
The application of the previously introduced concepts from information theory and statistical mechanics to complex networks, offers many intriguing possibilities, the most important one being the quantification of the amount of shared information between graphs and model fitting.
The relative entropy $S(\brho \| \bsigma)$ between two density matrices $\brho$ and $\bsigma$ is a nonnegative quantity that measures the expected amount of information lost when $\bsigma$ is used instead of $\brho$~\cite{dedomenico2016b,wilde2013} and it is defined as:
\begin{equation}\label{eq:relative_entropy}
S(\brho \| \bsigma) = \Tr{\brho (\log \brho - \log \bsigma)} \geq 0.
\end{equation}

From the linearity of the trace operator, it is apparent that~\Cref{eq:relative_entropy} consists of two terms.
The first term is the negative value of entropy of the empirical density $\brho$.
The second term $\log \mathcal{L} = -\Tr{\brho \log \bsigma}$ can be seen as the expected log-likelihood ratio between densities $\brho$ and $\bsigma$~\cite{wilde2013,dedomenico2016b,cover2006}.
For this reason we can also express~\Cref{eq:relative_entropy} as:
{\begin{equation}\label{eq:relative_entropysplit}
{S}(\brho \| \bsigma) = -S(\brho) - \log \mathcal{L(\brho,\bsigma)}.
\end{equation}}
Indeed~\Cref{eq:relative_entropysplit} can be thought as a measure to quantify the discrepancy between the density matrix $\brho$ of an observed network and the model density matrix.

In these settings, model optimization corresponds to finding the optimal parameters $\hat{\btheta}$ via minimization of the expectation of the relative entropy over all graphs with parameters $\btheta$:
\begin{align}\label{eq:minimizationproblem}
\hat{\btheta} = \underset{\btheta}{\argmin}\quad \mathbb{E}_{\btheta}\lbrack S(\brho\| \bsigma(\btheta) \rbrack.
\end{align}
Rigorous calculation of the expected relative entropy requires the knowledge of the spectral properties of the model Laplacian. These can be obtained via application of random matrix theory or by Monte Carlo sampling~\cite{nadakuditi2012,peixoto2013,nadakuditi2013}.
However in this case a continuous gradient based optimization cannot be applied as the relative entropy is no longer differentiable. Other methods, like simulated annealing or stochastic optimization~\cite{robbins1951,kiefer1952} can be applied in this case, with substantial computational burden.
Here, for simplicity we use a different approach.
As any random graph model depends on some parameters $\btheta = \{\theta_1,\ldots,\theta_k\}$, the model Laplacian is a matrix of random variables, where its elements are drawn from some distribution with parameters $\btheta$.
For example, in the Erd\H{o}s-R{\'{e}}nyi random graph, the Laplacian off-diagonal elements are Bernoulli random variables with expectation $-p$, while the diagonal elements are binomial random variables with expectation $(n-1)p$.
Hence, we denote the expectation operator of the model Laplacian $\bL(\btheta)$ at fixed parameters $\btheta$ as $\mathbb{E}_{\btheta}\lbrack \bL(\btheta) \rbrack$.

As discussed in the appendix ~\ref{app:expectedrelentropy} we approximate the expected relative entropy of Eq.~\ref{eq:minimizationproblem} with the relative entropy between the observed density and the density of the expected model Laplacian:
\begin{equation}\label{eq:approximationexpectedrelativeentropy}
\mathbb{E}_{\btheta}\lbrack S(\brho\| \bsigma(\btheta) ) \rbrack \approx S(\brho \| \bsigma(\mathbb{E}_{\btheta}[\bL(\btheta)])).
\end{equation}
The accuracy of the approximation in Eq.~\ref{eq:approximationexpectedrelativeentropy} is higher for large networks with low sparsity.
Finally, as the expected Laplacian $\mathbb{E}_{\btheta}[\bL]$ depends continuously on the parameters $\btheta$, we can use continuous optimization methods based on the analytically computed gradients with components (see~\Cref{app:gradientscalculation} for detailed calculation):
\begin{equation}\label{eq:gradloglikelihoodfinal}
\frac{\partial {S(\brho \| \bsigma(\mathbb{E}_{\btheta}[\bL])} }{\partial \theta_k} = \beta \Tr { \bigl( \brho-\bsigma(\mathbb{E}_{\btheta}[\bL(\btheta)]) \bigr) \frac{\partial \mathbb{E}_{\btheta}[\bL(\btheta)] }{\partial \theta_k} }.
\end{equation}
This last equation is the basis of the following sections and enables application of gradient based optimization methods.

\subsection{Thermodynamic interpretation}
The Klein inequality states that the quantum relative entropy of two density matrices is always non negative, and zero only in the case $\brho=\bsigma$~\cite{wilde2013}.
It is interesting to rework the expression for the 
relative entropy by making use of thermodynamic quantities~\cite{merhav2010,parrondo2015}.
For notational clarity here we set $\bL_{\bsigma}:=\mathbb{E}_{\btheta}[\bL(\btheta)]$.
We denote the Helmholtz free energies $F_\rho,F_\sigma$ of the two systems described by densities $\brho$ and $\bsigma(\mathbb{E}_{\btheta}[\bL(\btheta)])$ as:
\begin{align}
F_{\brho} &= -\beta^{-1}\log Z_{\brho}\nonumber \\
F_{\bsigma} &= -\beta^{-1}\log Z(\btheta) \nonumber
\end{align}
The partition functions are computed as $Z_{\brho} = \Tr{e^{-\beta \bL_{\brho}}}$, $Z_{\bsigma} = \Tr{e^{-\beta \bL_{\bsigma}}}$.
The ensemble averages of the empirical and model Laplacians are $\langle \bL_{\brho} \rangle_{\brho} = \Tr{\brho \bL_{\brho} }$ and $\langle \bL_{\bsigma} \rangle_{\brho}=\Tr{\brho \bL_{\bsigma} }$, where $\langle \cdot \rangle_{\brho}$ indicates thermal averaging with respect to the canonical distribution pertaining to the Laplacian of the observed network $\bL_{\brho}$.
After rearrangement of the terms, the expression for the relative entropy described in~\Cref{eq:relative_entropy} becomes:
{\begin{equation}\label{eq:thermodynamicdkl}
{S}(\brho \| \bsigma ) = \beta \left[ \left(F_{\brho} - F_{\bsigma} \right) - \left( \langle \bL_{\brho} \rangle_{\brho} - \langle \bL_{\bsigma} \rangle_{\brho} \right) \right] \geq 0.
\end{equation}
This expression is indeed general for any two density matrices and not only in the settings discussed above.
}
Clearly, the Klein inequality implies the following condition, also known as Gibbs' inequality in statistical physics~\cite{merhav2010}:
\begin{equation}\label{eq:gibbsinequality}
\langle \bL_{\bsigma} \rangle_{\brho} - F_{\bsigma} \geq \langle \bL_{\brho} \rangle_{\brho} - F_{\brho}.
\end{equation}
As the right hand side is independent on the parameters $\btheta$, the minimum relative entropy is obtained by direct minimization of the left-hand side of~\Cref{eq:gibbsinequality}.

This expression has a profound physical interpretation~\cite{merhav2010,deffner2010,parrondo2015}.
Let us consider a system with Hamiltonian $\bL_{\bsigma}(\btheta^*)$ where the parameters $\btheta^*$ are fixed.
This system is in equilibrium at temperature $1/\beta$ with a heat bath and its density matrix is $\bsigma^*:=\bsigma(\btheta^*)$.
Suppose we are given a sampling procedure to create real networks from the system described by the density $\bsigma^*$.
Naturally, the properties of a sampled network will slightly deviate from its ensemble average.
Indeed, except for a few trivial cases, the density $\brho$ of a single network sampled from $\bsigma^*$ will never be perfectly equal to $\bsigma^*$.

Thus, the generation of a random graph instance given a model described by Hamiltonian $\bL_{\bsigma}$ can be interpreted as a sudden perturbation where the Hamiltonian of the system is driven from $\bL_{\bsigma}$ to $\bL_{\brho}$.
In the general case this corresponds to an irreversible transformation, except for graphs where there are no possible rewirings preserving the given constraints, up to node permutations.

In this sense, minimization of the relative entropy is equivalent to finding a set of parameters $\tilde{\btheta}$ such that the work required to bring the system described by $\brho$ to a state $\bsigma(\tilde{\btheta})$ is minimum.
However, $\tilde{\btheta}$ is not guaranteed to be equal to $\btheta^*$ due to irreversibility of the sampling.
In order to precisely reconstruct the parameters $\tilde{\btheta} = \btheta^*$, minimization of relative entropy averaged over the whole set of samples is therefore needed.

The inverse temperature $\beta$ plays the role of a resolution parameter allowing one to compare two networks at different scales~\cite{dedomenico2016b,biamonte2017}.
Therefore, for maximum entropy models with linear constraints, the optimal $\beta$ tends to zero, as we are comparing the lowest order properties of the networks, linearly dependent on the adjacency matrix.
On the other hand, for $\beta$ tending to zero, the two density matrices $\brho$ and $\bsigma$ tend to identity, so any choice of the parameters trivially yields zero relative entropy.
Hence, to guide the optimization towards a non-trivial solution, one must start with some initial guess of $\beta_0$, isothermically find a local minimum, and afterwards decrease $\beta$.
Eventually, as $\beta$ tends to zero, the optimal solution will change slowly while making the relative entropy as small as possible.
Here, for the Erd\H{o}s-R{\'{e}}nyi and the planted partition model, we show analytically that correct reconstruction of the empirical density parameters is possible only in the limit $\beta \to 0$.

\subsection{Erd\H{o}s-R{\'{e}}nyi random graph}
The Erd\H{o}s-R{\'{e}}nyi random graph is the simplest example of random graph model~\cite{erdos1959}.
Each pair of nodes is connected by a link with constant probability $p$. Hence, the expected adjacency matrix and the Laplacian can be written as ${\mathbb{E}}[\bA_{\bsigma}]=p(\mathbf{1}-\bI)$ and ${\mathbb{E}}[\bL_{\bsigma}]=(n-1)p\bI - p(\mathbf{1}-\bI)$, where $\mathbf{1}$ is the $n\times n$ matrix of ones and $\bI$ is the identity matrix.
In this case it is possible to analytically find the optimum solution for the problem of relative entropy minimization. 
The partition function $Z_{\bsigma}(n,p)$ and the
{ ensemble average of the expected Laplacian} $\Tr{\brho \mathbb{E}[\bL]}$ of the model are:{
\begin{align}
Z_{\bsigma}(p) = (n-1)e^{-n \beta p} + 1 \nonumber \\
\Tr{{\mathbb{E}}[\bL(p)] \brho} = p(n - R(n,\beta)),
\end{align}}
where $R(n,\beta)=\Tr{\mathbf{1} \brho}=\sum_{i,j}^{n} \rho_{ij}$ is the grand sum of density matrix.
Both $Z_{\brho}$ and $F_{\brho}$ are observation dependent quantities and must be evaluated numerically from the observed network.
Finding the minimum of the left hand side of~\Cref{eq:gibbsinequality} corresponds to setting to zero its derivative with respect to $p$:
\begin{align}
&\frac{\partial}{\partial p}\left( \Tr{{{\mathbb{E}}}[\bL(p)] \brho} + \frac{\log Z_{\bsigma}(p)}{\beta} \right) = 0 \nonumber\\
&=\Tr{\frac{\partial {{\mathbb{E}}}[\bL(p)]}{\partial p} \brho} - \frac{n(n-1)}{e^{\beta n p}+(n-1)}=0
\end{align}
Solving for $p$ we can find an analytical expression for the optimal density $\tilde{p}$ that can be reconstructed by the model:
\begin{equation}
\tilde{p} = \frac{1}{n \beta}\log\left[ \frac{R(n,\beta)(n - 1)}{(n - R(n,\beta))} \right].
\end{equation}

The reconstruction of the observed empirical density $p^*=2m^*/(n(n-1))$ is only possible in the limit $\beta \to 0$.
It can be shown, with the help of computer algebra system, that:
\begin{equation}
\lim \limits_{\beta \to 0} \tilde{p} = p^* = \frac{2m^*}{n(n-1)}.
\end{equation}
We also extended the same calculations to the planted-partition model, the simplest extension of the Erd\H{o}s-R{\'{e}}nyi model to networks presenting a community structure~\cite{condon2000}.

\subsection{Planted partition model}
In the planted partition model, the nodal block membership vector {$c_i \in {\{0,\ldots,b \}}$} specifies to which one of the $b$ blocks the node $i$ belongs and it has the role of a hyper-parameter.
The model parameters are the intrablock and interblock link densities $\pin$ and $\pout$.
The expected adjacency matrix $\mathbb{E}[\bA]$ and Laplacian $\mathbb{E}[\bL]$ are:
\begin{align}
\mathbb{E}[\bA] &= \boldsymbol \delta \pin + (\mathbf{1}-\boldsymbol \delta)\pout \\
\mathbb{E}[\bL] &= \bI\left(\pin(n/b-1)+\pout(n/b)(b-1)\right) \nonumber \\&- \boldsymbol \delta \pin - (\mathbf{1}-\boldsymbol \delta)\pout,
\end{align}
where $\boldsymbol \delta = \{\delta_{c_i,c_j}\}$ is the block assignment matrix.
Analogous calculations as in the Erd\H{o}s-R{\'{e}}nyi case can be analytically performed in the planted partition model with exactly $b$ blocks of the same size.
In this case the partition function of the model $Z_{\bsigma}(\pin,\pout)$ becomes:
\begin{align}
Z_{\bsigma}&(\pin,\pout)\nonumber = \Tr{e^{-\beta {\mathbb{E}}[\bL_{\bsigma}]}} =\\=& \left(n-b\right)\exp{\left[-\beta n\left( p_{in} + (b-1)p_{out} \right)/b\right] } + \nonumber \\& + (b-1) \exp{\left[-\beta n p_{out}\right]} + 1.
\end{align}
Setting the gradients of relative entropy~(\ref{eq:gradloglikelihoodfinal}) to zero, results in a system of two equations:
\begin{align}
\begin{dcases}
\Tr{\frac{\partial {{\mathbb{E}}}[\bL_{\bsigma}(\pin,\pout)]}{\partial \pin} \brho } + \frac{1}{\beta }\frac{\partial \log Z_{\bsigma}(\pin,\pout)}{\partial \pin} = 0 \\ 
\Tr{\frac{\partial {{\mathbb{E}}}[\bL_{\bsigma}(\pin,\pout)]}{\partial \pout} \brho } + \frac{1}{\beta}\frac{\partial \log Z_{\bsigma}(\pin,\pout)}{\partial \pout} = 0.
\end{dcases}
\end{align}
An analytical solution is possible for $b=2$ blocks:
\begin{align}
\tilde{p}_{\textrm{in}} =& \frac{1}{\beta n}\log \left[ \frac{(n-2)^2 R (2Q-R)}{(n-2Q)^2} \right] \\
\tilde{p}_{\textrm{out}} = & \frac{1}{\beta n}\log \left[ \frac{R}{2Q-R} \right]
\end{align}
where $R$ is the grand-sum of $\brho$ and $Q=\Tr{\boldsymbol \delta \brho}$.
{As in the previous case, }the empirical intra-block and inter-block densities $\pin^*$ and $\pout^*$ can be reconstructed only in the limit of infinite temperature:
\begin{align}
\lim \limits_{\beta \to 0} \tilde{p}_{\textrm{in}} &= \pin^* \nonumber \\
\lim \limits_{\beta \to 0} \tilde{p}_{\textrm{out}} &= \pout^*.
\end{align}

Unfortunately, though, the calculations performed for these two last examples cannot be straightforwardly extended to the configuration model (UBCM) and other more complex variants of the exponential random graph model, as the expression of the partition function $Z_{\bsigma}$ for general models is too complex for a fully analytical treatment.
Therefore, we rely on numerical simulations to show that the limit $\beta \to 0$ yields a correct reconstruction of model parameters for the UBCM.

\subsection{Configuration model}
In the Undirected binary configuration model (UBCM) the model parameters are the hidden variables $\mathbf{x}=\{x_i\}$.
Given some network with fixed degree sequence, we can consider its Laplacian $\bL_{\brho}$ as the Laplacian of a graph sampled from the UBCM ensemble.

Importantly, being $\bL_{\brho}$ a realization of a random graph ensemble, we cannot expect to be able to reconstruct exactly the parameters $\mathbf{x}^*$ which generated that network.
As a demonstration of this concept we generated a random network with $n=40$ nodes and a degree sequence sampled from the uniform distribution.

Starting from a random solution $\mathbf{x}(t_0,\beta_0)$, we minimized the relative entropy at an initial guess of $\beta_0 \gg 1$ to obtain the new solution $\mathbf{x}(t_1, \beta_0)$.
This procedure was repeated gradually decreasing $\beta$ at each iteration, until the solution was not changing considerably, for $\beta$ close to $0$.

As a measure of convergence, we chose the difference of the total number of links, which in thermodynamic interpretation equals to half the difference of the total energies: $\Delta m = (\Tr{{\mathbb{E}}[\bL_{\bsigma}]} - \Tr{\bL_{\brho}})/2$.
The bigger the absolute value of $\Delta m$, the less similar the empirical network is from the average realization of the ensemble.

\begin{figure*}
\centering
\includegraphics[width=1.0\textwidth]{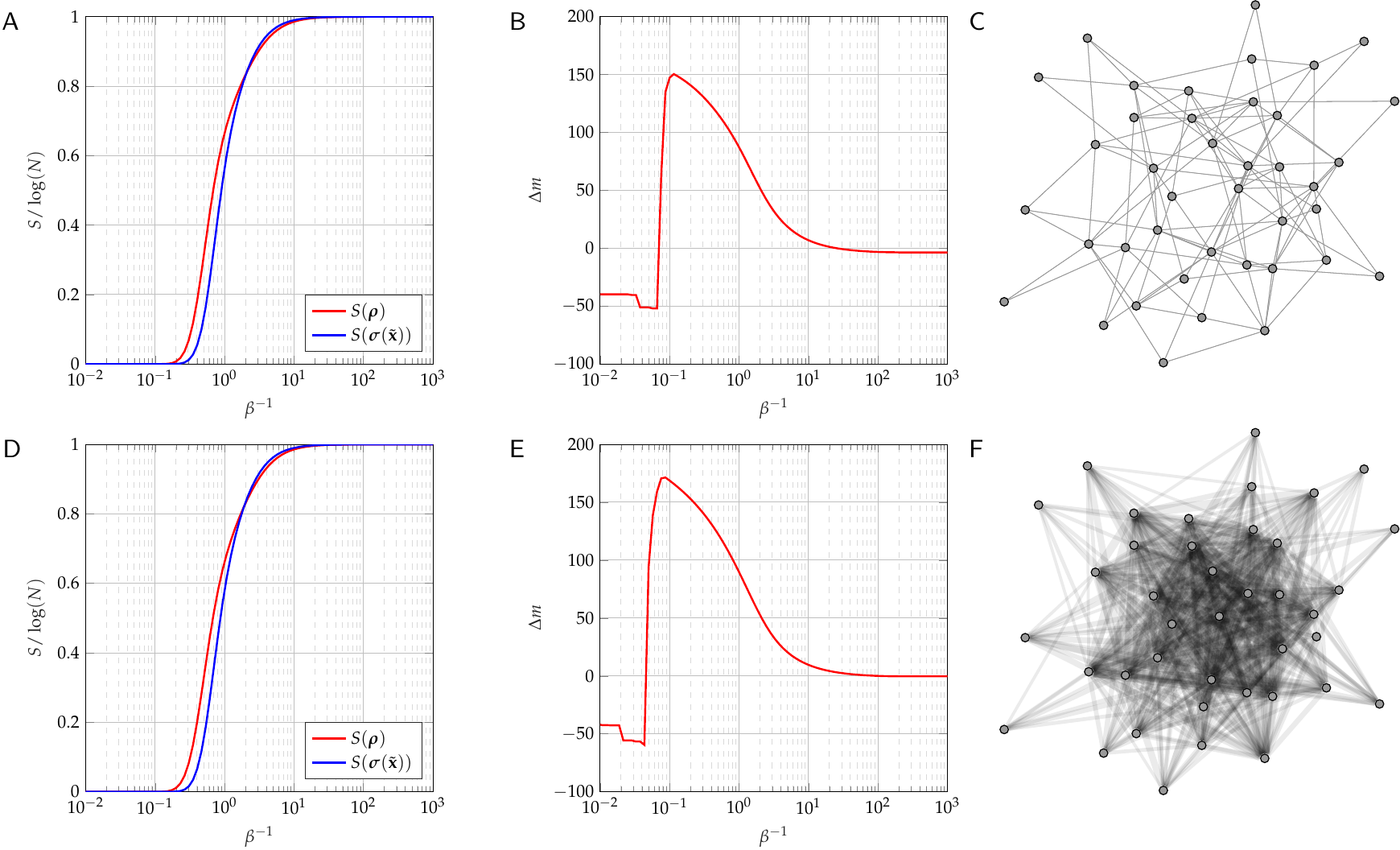}
\caption{Normalized von Neumann entropy and reconstruction error of a random network (top panels) and its ensemble average with the same degree sequence (bottom panels). In panels A,D the red curves are the spectral entropy of the observed network described by density $\brho$, blue curves are the spectral entropies of the model network at optimum parameters $\bsigma(\tilde{\btheta})$. In panel B a systematic error exists at every $\beta$ (even in the limit $\beta \to 0$): {in the spectral framework }the UBCM cannot fit all the specific properties of the highly ordered network. On the other hand, an ensemble average network has a structure that can be better reconstructed by the UBCM, as shown by the error $\Delta m$ going to zero in Panel E.}
\label{fig:ubcmrandomdegree}
\end{figure*}

In~\Cref{fig:ubcmrandomdegree}A we plotted the spectral entropies of the empirical network (red line) and the fitted model at optimal solution $\tilde{\mathbf{x}}$ (blue line) as a function of $\beta$. 
\Cref{fig:ubcmrandomdegree}B shows the difference in the number of links $\Delta m$ as a function of $\beta$.
At the optimal solution for $\beta \to 0$, there is a small deviation in total number of links ($\Delta m \approx -3.64$).
This is explained by the irreversibility of the sampling process, that implies inability to precisely reconstruct the true parameters $\mathbf{x}^*$ from only one sample.
However, the deviation of the reconstructed parameters $\tilde{\mathbf{x}}$ from $\mathbf{x}^*$ can be reduced with enough samples of the random graph ensemble, as shown in~\Cref{fig:ubcmrandomdegree} (panels C and D).

\begin{figure*}
\centering
\includegraphics[width=1.0\textwidth]{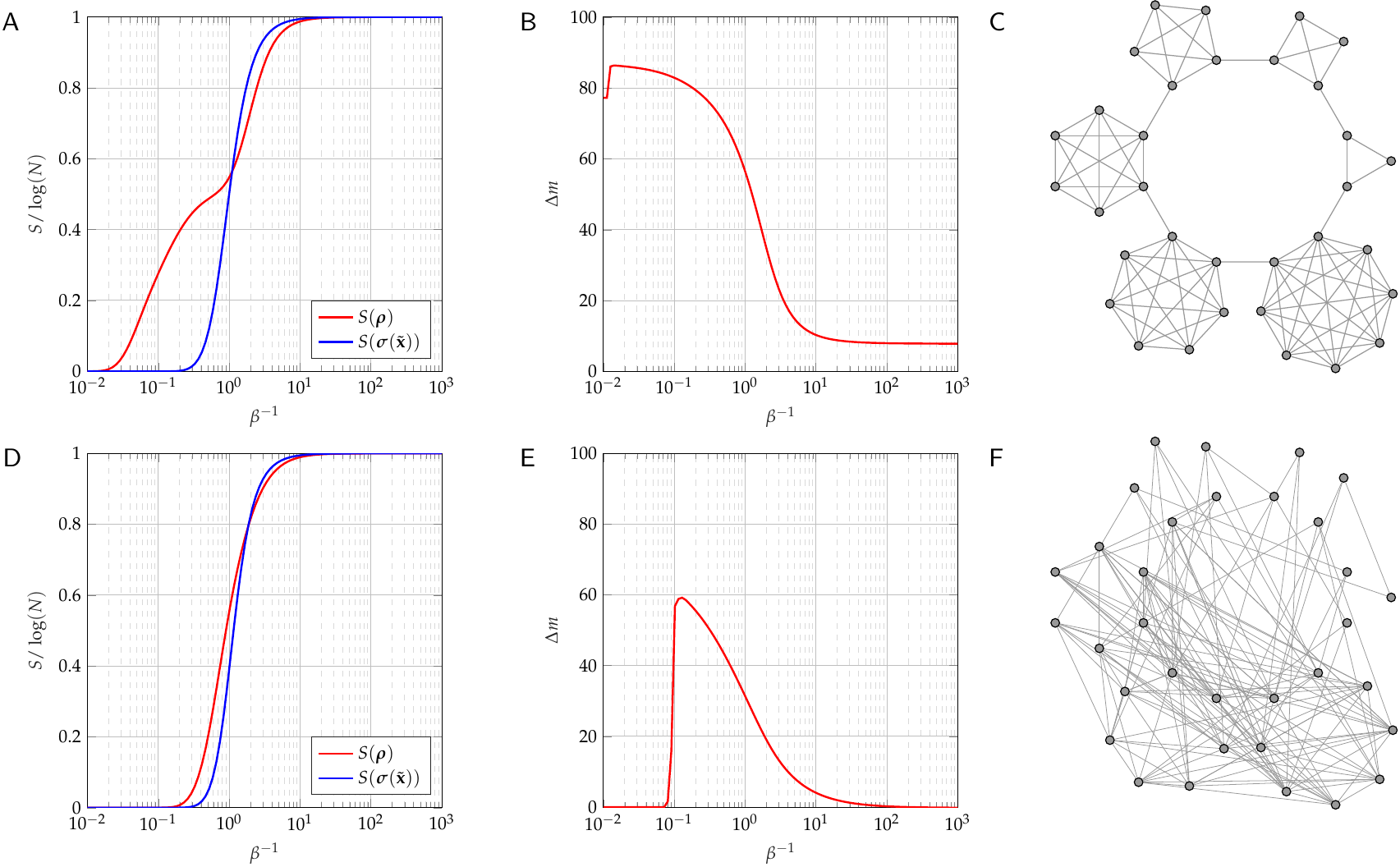}
\caption{Normalized von Neumann entropy and reconstruction error of two networks: a ring of cliques (top panels) and its degree-preserving random rewiring (bottom panels).
In panels A,D the red curves are the spectral entropy of the observed network described by density $\brho$, blue curves are the spectral entropies of the model network at optimum parameters $\bsigma(\tilde{\btheta})$. In panel B a systematic error exists at every $\beta$ (even in the limit $\beta \to 0$): the UBCM cannot fit all the specific properties of the highly ordered network.
On the other hand, a degree preserving rewired network has a structure that can be better reconstructed by the UBCM, as shown by the error $\Delta m$ going to zero in Panel E.}
\label{fig:ubcmringcliquefit}
\end{figure*}

As a second example we chose a toy network consisting of a number of cliques of increasing size connected in a ring, and one of its degree-preserving random rewiring (~\Cref{fig:ubcmringcliquefit}C,F).
In the ordered case, the clique structure cannot be accounted by a first-order average model alone, making that specific instance highly unlikely {in our framework} when sampling from the configuration model.
Therefore, following the fitting procedure described above, one can see a significant difference in the number of links between model and data $\Delta m$ even at a very small beta~\Cref{fig:ubcmringcliquefit}B.
In other words, in the ordered case, the degree sequence alone cannot explain the differences in the spectral entropies, thus indicating the presence of genuine regular patterns that substantially alter the properties of diffusion of the random walk{er} defined by the density matrix.
Indeed, this difference reflects the intrinsic inability of the model to account for the characteristic structure of the underlying network.

After fitting, non vanishing $\Delta m$ serves as an indicator of the presence of ordered patterns in the given network that are not explained by this model alone.
To test this idea we applied the same optimization technique to the degree-preserving randomly rewired network in Figure~\ref{fig:ubcmringcliquefit}F, and plotted the results in~\Cref{fig:ubcmringcliquefit}D,E.
In this case the random rewiring made the empirical network more adherent to the optimal reconstruction by the model and the difference in the total number of links at $\beta$ close to zero is much smaller than in the ordered case.
This is also evident by the better adherence in the spectral entropies, as shown in~\Cref{fig:ubcmringcliquefit}D.

Importantly, the spectral entropy optimization framework described above can be applied to descriptive network models other than those described by the exponential random graph model.

\section{Spatial models}
The embedding of a network in a two or three dimensional space has bearings on 
its topological properties.
When the formation of links has a cost associated with distance, the model must accomodate additional spatial constraints, which introduce correlation between topological and geometrical organization~\cite{barthelemy2011}.
An example is represented by neural networks, in which communication between neurons implies a metabolic cost that depends on their distance~\cite{bullmore2012,betzel2017}.
The material and metabolic constraints of neuronal wiring are factors that contributed to shaping brain architecture~\cite{bullmore2012,stam2012,ribrault2011}.
Computational and empirical studies converged on the result that a multiscale organization of modules inside modules is the one that satisfies the constraints imposed by minimization of energetic cost and spatial embedding~\cite{bullmore2012,doucet2011,betzel2017,kaiser2006}.
Here, wiring cost includes the physical volume of axons and synapses, the energetic demand for signal transmission, additional processing cost for noise correction over long distance signaling and sustenance of the necessary neuroglia that support neuronal activities~\cite{bullmore2012}. 
Therefore, it is tempting to assume that the expected number of neural fibers between two areas could be expressed as a decreasing function of their length.
With this hypothesis in mind, we verified the ability of our {optimization }approach to work with a simple descriptive model of the observed neural connectivity in the macaque cortex~\cite{markov2014,ercsey-ravasz2013}.
The model, called Exponential Distance Rule (EDR),
is a dense weighted network model describing the decline in the expected number of axonal projections $w_{ij}$ as a function of the inter-areal distances $d_{ij}$ and a tunable decay parameter $\ell \in \mathbb{R}$:
\begin{equation}
{\mathbb{E}}[w_{ij}] = C e^{-\ell d_{ij}},
\end{equation}
where $C$ is a normalization constant. Here, the distances $d_{ij}$ are measured along the shortest path connecting areas via white matter, approximating the axonal distance~\cite{markov2014}.

\begin{figure*}[tb]
\centering
\includegraphics[width=1.0\textwidth]{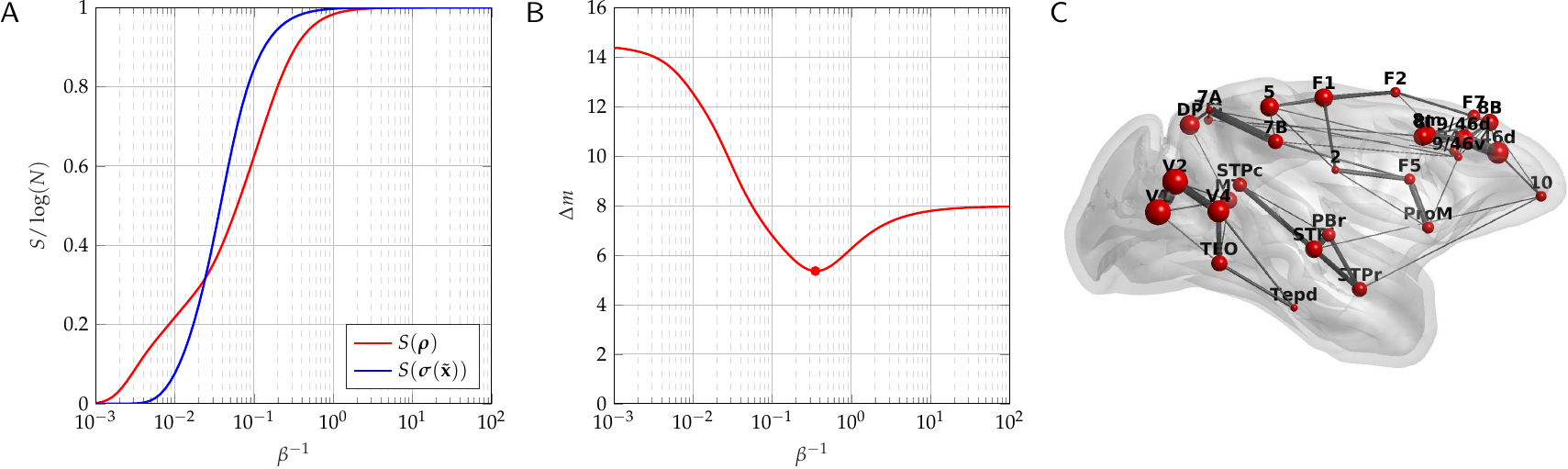}
\caption{Normalized von Neumann entropy and reconstruction error of the macaque brain connectivity network (panel C). In panel A the red curve is the spectral entropy of the observed network described by density $\brho$, blue curve is the spectral entropy of the model network at optimum parameter $\bsigma(\tilde{\btheta})$.
The optimal hyperparameter $\beta$ is found in panel B where the reconstruction error $\Delta m$ achieve its minimum denoted by a red dot.}
\label{fig:edrmacaque}
\end{figure*}

We used a dataset of cortico-cortical connectivity generated from retrograde tracing experiments in the macaque brain~\cite{ercsey-ravasz2013,markov2014}.
Following the procedure introduced in the previous sections, we fitted the macaque connectome network with the EDR model.
Differently from the random graph models described in previous sections, we found an optimal inverse-temperature parameter $\beta$ that minimizes the reconstruction error, as shown in~\Cref{fig:edrmacaque}B.
At this optimal $\beta^* \approx 3.05$ the reconstructed decay parameter $\tilde{\ell} = 0.1505$ $[\textrm{mm}]^{-1}$ is comparable to the values obtained by the three methods applied in the original paper from Ref~\cite{ercsey-ravasz2013}.
The non-vanishing difference in total weights between reconstructed and object networks indicates that the model cannot account completely for the structures such as the high density core observed in the connectome~\cite{ercsey-ravasz2013}.
Hence, a non-zero optimal value for $\beta$ suggests the existence of a scale at which the model best describes the topological properties of the network.

\section{Conclusion}
The spectral entropies framework enables comparing networks taking into account the whole structure at multiple scales.
However, this approach introduces a hyperparameter $\beta$ that plays the role of an inverse temperature, and whose tuning is critical for the correct estimate of the model parameters. 

Leveraging a thermodynamic analogy, we have shown that the optimal value of the hyperparameter is model dependent and reflects the scales at which the model best describes the empirical network.
Moreover, we have described procedures to determine $\beta$ for the model parameter optimization and for a tractable approximation of the expected relative entropy.

Specifically, we focused on three examples from the exponential random graph model, namely the Erd\H{o}s-R{\'{e}}nyi, a planted partition and undirected binary configuration model.
In the Erd\H{o}s-R{\'{e}}nyi model and in the planted partition model, we analytically demonstrated that correct reconstruction is possible only in the infinite temperature limit.
In the configuration model this hypothesis was verified numerically. 
The presence of ordered structures unaccounted for by the model is reflected in a bias of the total energy, corresponding to the total number of links in the reconstructed network.

Motivated by these findings in synthetic networks, we applied the spectral entropy framework to a real-world network of the macaque brain structural connectome.
A structural connectome is a spatial network whose development is thought to be constrained by geometrical and wiring cost factors. 
Hence, we evaluated an exponential distance rule model that assumes that the weight of inter-areal connections is a decreasing function of distance.
We demonstrate the existence of a non-zero optimal value of $\beta$ for the computation of model parameters.
However, the residual energy bias indicates the network structure at certain scales cannot be described by the exponential distance rule model alone. 

The procedures demonstrated here make it possible to use relative entropy methods for practical applications to the study of models of real-world networks, effectively realizing a conceptual step from classical maximum likelihood methods to their density matrix based counterparts.

\section*{Acknowledgments}
This project has received funding from the European Union's Horizon 2020 Research and Innovation Program under grant agreement No 668863.

\appendix
\section{Approximation of expected relative entropy}\label{app:expectedrelentropy}
We can exploit the commutativity and linearity of the trace and expectation operators to obtain a simpler expression for the expected relative entropy:
\begin{align}\label{eq:expected_relative_entropy}
\mathbb{E}_{\btheta}\lbrack S(\brho\| \bsigma(\btheta) \rbrack = \Tr{\brho \log \brho} - \Tr{\brho \mathbb{E}_{\btheta}\left \lbrack \log \bsigma (\btheta) \right \rbrack}.
\end{align}
By the positive-definiteness of the density $\bsigma$, we have:
\begin{equation}\label{eq:applogsigma}
\log (\bsigma(\btheta)) = -\beta \bL(\btheta) - \mathbf{I}\log Z(\btheta).
\end{equation}
Plugging this into the expression of the expected relative entropy we obtain:
\begin{align}
\mathbb{E}_{\btheta}\lbrack S(\brho\| \bsigma(\btheta) \rbrack = \Tr{\brho \left( \log \brho + \beta \mathbb{E}_{\btheta}[\bL(\btheta)] + \bI\mathbb{E}_{\btheta}\left[\log Z(\btheta)\right] \right)}
\end{align}
This last expression depends on the expected Laplacian of the model $\mathbb{E}_{\btheta}\lbrack \bL(\btheta)\rbrack$ and on the expected log-partition function $\mathbb{E}_{\btheta}[\log Z(\btheta)]$.
An analytical estimate of the expected Laplacian as function of the parameters $\btheta$ can readily be obtained, but computation of the expected log-partition function $\mathbb{E}_{\btheta}[\log Z(\btheta)]$ is more difficult and requires techniques from random matrix theory~\cite{nadakuditi2012,peixoto2013,nadakuditi2013}.
This is clear as the trace of a matrix is equal to the sum of its eigenvalues, yielding:
\begin{equation}\label{eq:expectedz}
\mathbb{E}_{\btheta}\left[\log Z(\btheta)\right] = \mathbb{E}_{\btheta}\left[\log\left( \sum_{i=1}^n e^{-\beta \lambda_i(\bL(\btheta))} \right)\right].
\end{equation}
We can estimate the expected log-partition function by means of matrix concentration arguments~\cite{tropp2015,cape2017}.
A random matrix is said to \emph{concentrate} when, given some spectral norm, one can tightly bound the spectral norm of the difference from its expected value~\cite{qiu2014}.
In the case of network-related matrices, the eigenvalues of the Laplacian and those of its expectation are strictly related and can be tightly bounded with high probability~\cite{oliveira2009,preciado2017,cape2017}.
This approximation becomes more precise, the larger and denser the graphs are~\cite{le2015,le2017}, an effect of the concentration of measure phenomenon.
Therefore, following the ideas presented in references~\cite{cape2017,oliveira2009} that apply in our same settings, we replace $\lambda_i(\bL(\btheta))$ with their counterparts from the expected Laplacian $\lambda_i(\mathbb{E}_{\btheta}[\bL])$.
Substituting back, we recover an expression that involves the relative entropy between the observed density and the density of the expected Laplacian:
\begin{equation}
\mathbb{E}_{\btheta}\lbrack S(\brho\| \bsigma(\btheta) ) \rbrack \approx S(\brho \| \bsigma(\mathbb{E}_{\btheta}[\bL])).
\end{equation}

\section{Gradients of the relative entropy}\label{app:gradientscalculation}
Here we present the analytical calculation of the gradients of the relative entropy described in Eq.~\ref{eq:gradloglikelihoodfinal}.
We can decompose the relative entropy using Eq.~\ref{eq:applogsigma} as:
\begin{align}\label{eq:loglikelihood2}
S(\brho\| \bsigma(\mathbb{E}_{\btheta}[\bL])) = 
\beta \Tr{\brho \mathbb{E}_{\btheta}[\bL(\btheta)]} + \log \Tr{e^{-\beta \mathbb{E}_{\btheta}[\bL(\btheta)]}}
\end{align}
where we have used the fact that $\Tr{\brho \bI}=\Tr{\brho}=1$ by definition of density matrix.
Taking the derivatives with respect to the $k$-th parameter $\theta_k$, and by linearity of the trace operator, we get:
\begin{align}\label{eq:gradloglikelihood}
\frac{\partial S(\brho \| \bsigma(\mathbb{E}_{\btheta}[\bL(\btheta)])) }{\partial \theta_k} = & \beta \Tr{\brho \frac{\partial \mathbb{E}_{\btheta}\lbrack \bL(\btheta) \rbrack}{\partial \theta_k}}\nonumber \\ &+ \frac{\partial }{\partial \theta_k} \log \Tr{e^{-\beta \mathbb{E}_{\btheta}[\bL(\btheta)]}}
\end{align}
The following identity holds for the derivatives of the matrix exponential function:
\begin{equation}\label{eq:exponentialmapderiv}
\frac{d}{d t}\Tr{ e^{\mathbf{X}(t)} } = \Tr{ e^{\mathbf{X}(t)} \frac{d \mathbf{X}(t)}{d t} },
\end{equation}
so we can simply compute the second term involving the log-trace by standard calculus tools.
After some algebraic manipulation, we finally arrive to the expression for the derivative of the relative entropy with respect to the model parameters as described in the main text:
\begin{equation}
\frac{\partial S(\brho \| \bsigma(\mathbb{E}[\bL]))}{\partial \theta_k} = \beta \Tr{ \left( \brho - \bsigma(\mathbb{E}_{\btheta}[\bL(\btheta)]) \right)\frac{\partial \mathbb{E}_{\btheta}[\bL(\btheta)]}{\partial \theta_k}}
\end{equation}


\begin{thebibliography}{60}
\providecommand{\natexlab}[1]{#1}
\providecommand{\url}[1]{\texttt{#1}}
\expandafter\ifx\csname urlstyle\endcsname\relax
  \providecommand{\doi}[1]{doi: #1}\else
  \providecommand{\doi}{doi: \begingroup \urlstyle{rm}\Url}\fi

\bibitem[Newman(2010)]{newman2010book}
Mark Newman.
\newblock \emph{Networks: An Introduction}.
\newblock OUP Oxford, 2010.

\bibitem[Barabasi and Albert(1999)]{barabasi1999}
Albert-Laszlo Barabasi and Reka Albert.
\newblock Emergence of scaling in random networks.
\newblock \emph{Science}, 286\penalty0 (5439):\penalty0 509--512, 1999.

\bibitem[Caldarelli(2007)]{caldarelli2007}
Guido Caldarelli.
\newblock \emph{Scale-free networks: complex webs in nature and technology}.
\newblock Oxford University Press, 2007.

\bibitem[Bullmore and Sporns(2009)]{bullmore2009}
Ed~Bullmore and Olaf Sporns.
\newblock {Complex brain networks: graph theoretical analysis of structural and
  functional systems.}
\newblock \emph{Nat. Rev. Neurosci.}, 10\penalty0 (3):\penalty0 186--198, 2009.
\newblock ISSN 1471-0048.

\bibitem[Wasserman and Faust(1994)]{wasserman1994}
Stanley Wasserman and Katherine Faust.
\newblock \emph{Social network analysis: Methods and applications}, volume~8.
\newblock Cambridge university press, 1994.

\bibitem[Squartini and Garlaschelli(2017)]{squartini2017book}
Tiziano Squartini and Diego Garlaschelli.
\newblock \emph{Maximum-Entropy Networks: Pattern Detection, Network
  Reconstruction and Graph Combinatorics}.
\newblock Springer, 2017.

\bibitem[{De Domenico} and Biamonte(2016)]{dedomenico2016b}
Manlio {De Domenico} and Jacob Biamonte.
\newblock {Spectral entropies as information-theoretic tools for complex
  network comparison}.
\newblock \emph{Phys. Rev. X}, 041062:\penalty0 1--13, 2016.

\bibitem[Parrondo et~al.(2015)Parrondo, Horowitz, and Sagawa]{parrondo2015}
Juan~M.R. Parrondo, Jordan~M. Horowitz, and Takahiro Sagawa.
\newblock {Thermodynamics of information}.
\newblock \emph{Nat. Phys.}, 11\penalty0 (2):\penalty0 131--139, 2015.

\bibitem[Park and Newman(2004)]{park2004}
Juyong Park and M.~E.~J. Newman.
\newblock {Statistical mechanics of networks}.
\newblock \emph{Phys. Rev. E}, 70\penalty0 (6):\penalty0 066117, 2004.

\bibitem[Jaynes(1957)]{jaynes1957b}
E.~T. Jaynes.
\newblock {Information theory and statistical mechanics}.
\newblock \emph{Phys. Rev.}, 106\penalty0 (4):\penalty0 620--630, 1957.

\bibitem[Erd{\"{o}}s and R{\'{e}}nyi(1959)]{erdos1959}
P~Erd{\"{o}}s and a~R{\'{e}}nyi.
\newblock {On random graphs}.
\newblock \emph{Publ. Math.}, 6:\penalty0 290--297, 1959.
\newblock ISSN 00029947.

\bibitem[Caldarelli et~al.(2002)Caldarelli, Capocci, De~Los~Rios, and
  Mu\~noz]{caldarelli2002}
G.~Caldarelli, A.~Capocci, P.~De~Los~Rios, and M.~A. Mu\~noz.
\newblock Scale-free networks from varying vertex intrinsic fitness.
\newblock \emph{Phys. Rev. Lett.}, 89:\penalty0 258702, 2002.

\bibitem[Squartini et~al.(2015)Squartini, Mastrandrea, and
  Garlaschelli]{squartini2015}
Tiziano Squartini, Rossana Mastrandrea, and Diego Garlaschelli.
\newblock {Unbiased sampling of network ensembles}.
\newblock \emph{New J. Phys.}, 17\penalty0 (2):\penalty0 023052, 2015.
\newblock ISSN 1367-2630.

\bibitem[Squartini and Garlaschelli(2014)]{squartini2014}
Tiziano Squartini and Diego Garlaschelli.
\newblock {Jan Tinbergen's legacy for economic networks: from the gravity model
  to quantum statistics}.
\newblock \emph{Econophysics of Agent-Based Models}, pages 161--186, 2014.

\bibitem[Garlaschelli and Loffredo(2008)]{garlaschelli2008}
Diego Garlaschelli and Maria~I. Loffredo.
\newblock {Maximum likelihood: Extracting unbiased information from complex
  networks}.
\newblock \emph{Phys. Rev. E}, 78\penalty0 (1):\penalty0 1--4, 2008.

\bibitem[Braunstein et~al.(2006)Braunstein, Ghosh, and
  Severini]{braunstein2006a}
Samuel~L. Braunstein, Sibasish Ghosh, and Simone Severini.
\newblock {The Laplacian of a graph as a density Matrix: A basic combinatorial
  approach to separability of mixed states}.
\newblock \emph{Ann. Comb.}, 10\penalty0 (3):\penalty0 291--317, 2006.

\bibitem[Estrada(2011)]{estrada2011}
Ernesto Estrada.
\newblock \emph{The Structure of Complex Networks: Theory and Applications}.
\newblock Oxford University Press, Inc., New York, NY, USA, 2011.

\bibitem[Anderson(1985)]{anderson1985}
William~N Anderson.
\newblock {Eigenvalues of the Laplacian of a graph}.
\newblock \emph{Linear Multilinear A.}, 18\penalty0 (2):\penalty0 141--145,
  1985.

\bibitem[Merris(1994)]{merris1994a}
Russell Merris.
\newblock {Laplacian matrices of graphs: a survey}.
\newblock \emph{Linear Algebra Appl.}, 197-198\penalty0 (C):\penalty0 143--176,
  1994.

\bibitem[de~Lange et~al.(2014)de~Lange, de~Reus, and van~den
  Heuvel]{delange2014}
Siemon~C. de~Lange, Marcel~A. de~Reus, and Martijn~P. van~den Heuvel.
\newblock {The Laplacian spectrum of neural networks}.
\newblock \emph{Frontiers in Computational Neuroscience}, 7\penalty0
  (January):\penalty0 1--12, 2014.
\newblock ISSN 1662-5188.

\bibitem[de~Lange et~al.(2016)de~Lange, van~den Heuvel, and
  de~Reus]{delange2016}
Siemon~C. de~Lange, Martijn~P. van~den Heuvel, and Marcel~A. de~Reus.
\newblock {The role of symmetry in neural networks and their Laplacian
  spectra}.
\newblock \emph{NeuroImage}, 141:\penalty0 357--365, 2016.
\newblock ISSN 10959572.

\bibitem[Cheeger(1970)]{cheeger1970}
Jeff Cheeger.
\newblock A lower bound for the smallest eigenvalue of the laplacian.
\newblock \emph{Problems in analysis}, pages 195--199, 1970.

\bibitem[Donetti et~al.(2006)Donetti, Neri, and Mu{\~{n}}oz]{donetti2006}
Luca Donetti, Franco Neri, and Miguel~A Mu{\~{n}}oz.
\newblock {Optimal network topologies: expanders, cages, Ramanujan graphs,
  entangled networks and all that}.
\newblock \emph{J. Stat. Mech-Theory E}, 2006\penalty0 (08):\penalty0 P08007,
  2006.

\bibitem[Lov{\'{a}}sz(1993)]{lovasz1993}
L~Lov{\'{a}}sz.
\newblock {Random walks on graphs: A survey}.
\newblock \emph{Bolyai Math. Stud.}, 2\penalty0 (Volume 2):\penalty0 1--46,
  1993.

\bibitem[Masuda et~al.(2017)Masuda, Porter, and Lambiotte]{masuda2017}
Naoki Masuda, Mason~A. Porter, and Renaud Lambiotte.
\newblock {Random walks and diffusion on networks}.
\newblock \emph{Phys. Rep.}, 2017.

\bibitem[Bray and Rodgers(1988)]{bray1988}
A.~J. Bray and G.~J. Rodgers.
\newblock {Diffusion in a sparsely connected space: A model for glassy
  relaxation}.
\newblock \emph{Phys. Rev. B}, 38\penalty0 (16):\penalty0 11461--11470, 1988.

\bibitem[Mohar et~al.(1991)Mohar, Alavi, Chartrand, and
  Oellermann]{mohar1991laplacian}
Bojan Mohar, Y~Alavi, G~Chartrand, and OR~Oellermann.
\newblock The laplacian spectrum of graphs.
\newblock \emph{Graph theory, combinatorics, and applications}, 2\penalty0
  (871-898):\penalty0 12, 1991.

\bibitem[Anand et~al.(2011)Anand, Bianconi, and Severini]{anand2011}
Kartik Anand, Ginestra Bianconi, and Simone Severini.
\newblock {Shannon and von Neumann entropy of random networks with
  heterogeneous expected degree}.
\newblock \emph{Phys. Rev. E}, 83\penalty0 (3):\penalty0 1--8, 2011.

\bibitem[Wilde(2013)]{wilde2013}
Mark~M Wilde.
\newblock \emph{Quantum information theory}.
\newblock Cambridge University Press, 2013.

\bibitem[Higham(2008)]{higham2008}
Nicholas~J Higham.
\newblock \emph{Functions of matrices: theory and computation}.
\newblock SIAM, 2008.

\bibitem[Jaynes(2003)]{jaynes2003}
E.~T. Jaynes.
\newblock {Probability Theory: The Logic of Science.}
\newblock \emph{The Mathematical Intelligencer}, 27\penalty0 (2):\penalty0
  83--83, 2003.
\newblock ISSN 0343-6993.

\bibitem[Biamonte et~al.(2017)Biamonte, Faccin, and {De
  Domenico}]{biamonte2017}
Jacob Biamonte, Mauro Faccin, and Manlio {De Domenico}.
\newblock {Complex Networks: from Classical to Quantum}.
\newblock \emph{arXiv preprint arXiv:1702.08459}, 2017.

\bibitem[Estrada and Hatano(2008)]{estrada2008}
Ernesto Estrada and Naomichi Hatano.
\newblock {Communicability in complex networks}.
\newblock \emph{Phys. Rev. E}, 77\penalty0 (3):\penalty0 1--12, 2008.

\bibitem[Faccin et~al.(2013)Faccin, Johnson, Biamonte, Kais, and
  Migda{\l}]{faccin2013}
Mauro Faccin, Tomi Johnson, Jacob Biamonte, Sabre Kais, and Piotr Migda{\l}.
\newblock {Degree Distribution in Quantum Walks on Complex Networks}.
\newblock \emph{Phys. Rev. X}, 3\penalty0 (4):\penalty0 041007, 2013.

\bibitem[Estrada et~al.(2012)Estrada, Hatano, and Benzi]{estrada2012a}
Ernesto Estrada, Naomichi Hatano, and Michele Benzi.
\newblock {The physics of communicability in complex networks}.
\newblock \emph{Phys. Rep.}, 514\penalty0 (3):\penalty0 89--119, 2012.

\bibitem[Cover and Thomas(2006)]{cover2006}
Thomas~M. Cover and Joy~A. Thomas.
\newblock \emph{Elements of Information Theory}.
\newblock Wiley-Interscience, 2006.

\bibitem[Nadakuditi and Newman(2012)]{nadakuditi2012}
Raj~Rao Nadakuditi and Mark~EJ Newman.
\newblock Graph spectra and the detectability of community structure in
  networks.
\newblock \emph{Phys. Rev. Lett.}, 108\penalty0 (18):\penalty0 188701, 2012.

\bibitem[Peixoto(2013)]{peixoto2013}
Tiago~P. Peixoto.
\newblock Eigenvalue spectra of modular networks.
\newblock \emph{Phys. Rev. Lett.}, 111:\penalty0 098701, Aug 2013.

\bibitem[Nadakuditi and Newman(2013)]{nadakuditi2013}
Raj~Rao Nadakuditi and Mark~EJ Newman.
\newblock Spectra of random graphs with arbitrary expected degrees.
\newblock \emph{Phys. Rev. E}, 87\penalty0 (1):\penalty0 012803, 2013.

\bibitem[Robbins and Monro(1951)]{robbins1951}
Herbert Robbins and Sutton Monro.
\newblock A stochastic approximation method.
\newblock \emph{The annals of mathematical statistics}, pages 400--407, 1951.

\bibitem[Kiefer and Wolfowitz(1952)]{kiefer1952}
Jack Kiefer and Jacob Wolfowitz.
\newblock Stochastic estimation of the maximum of a regression function.
\newblock \emph{The Annals of Mathematical Statistics}, pages 462--466, 1952.

\bibitem[Merhav(2010)]{merhav2010}
Neri Merhav.
\newblock {Statistical Physics and Information Theory}.
\newblock \emph{Foundations and Trends in Communications and Information
  Theory}, 6\penalty0 (1-2):\penalty0 1--212, 2010.

\bibitem[Deffner and Lutz(2010)]{deffner2010}
Sebastian Deffner and Eric Lutz.
\newblock {Generalized clausius inequality for nonequilibrium quantum
  processes}.
\newblock \emph{Phys. Rev. Lett.}, 105\penalty0 (17):\penalty0 1--4, 2010.

\bibitem[Condon and Karp(2000)]{condon2000}
Anne Condon and Richard~M Karp.
\newblock {on the Planted Partition Model}.
\newblock \emph{Electr. Eng.}, pages 116--140, 2000.

\bibitem[Barth{\'{e}}lemy(2011)]{barthelemy2011}
Marc Barth{\'{e}}lemy.
\newblock {Spatial networks}.
\newblock \emph{Phys. Rep.}, 499\penalty0 (1-3):\penalty0 1--101, 2011.

\bibitem[Bullmore and Sporns(2012)]{bullmore2012}
Ed~Bullmore and Olaf Sporns.
\newblock {The economy of brain network organization.}
\newblock \emph{Nat. Rev. Neurosci.}, 13\penalty0 (5):\penalty0 336--349, 2012.
\newblock ISSN 1471-0048.

\bibitem[Betzel and Bassett(2017)]{betzel2017}
Richard~F Betzel and Danielle~S Bassett.
\newblock Generative models for network neuroscience: prospects and promise.
\newblock \emph{J. R. Soc. Interface}, 14\penalty0 (136):\penalty0 20170623,
  2017.

\bibitem[Stam and van Straaten(2012)]{stam2012}
C.~J. Stam and E.~C~W van Straaten.
\newblock {The organization of physiological brain networks}.
\newblock \emph{Clin. Neurophysiol.}, 123\penalty0 (6):\penalty0 1067--1087,
  2012.
\newblock ISSN 13882457.

\bibitem[Ribrault et~al.(2011)Ribrault, Sekimoto, and Triller]{ribrault2011}
Claire Ribrault, Ken Sekimoto, and Antoine Triller.
\newblock {From the stochasticity of molecular processes to the variability of
  synaptic transmission}.
\newblock \emph{Nat. Rev. Neurosci.}, 12\penalty0 (7):\penalty0 375--387, 2011.

\bibitem[Doucet et~al.(2011)Doucet, Naveau, Petit, Delcroix, Zago, Crivello,
  Jobard, Tzourio-Mazoyer, Mazoyer, Mellet, and Joliot]{doucet2011}
Ga{\"{e}}lle Doucet, Mika{\"{e}}l Naveau, Laurent Petit, Nicolas Delcroix,
  Laure Zago, Fabrice Crivello, Ga{\"{e}}l Jobard, Nathalie Tzourio-Mazoyer,
  Bernard Mazoyer, Emmanuel Mellet, and Marc Joliot.
\newblock {Brain activity at rest: a multiscale hierarchical functional
  organization.}
\newblock \emph{J. Neurophysiol.}, 105\penalty0 (6):\penalty0 2753--2763, 2011.
\newblock ISSN 1522-1598.

\bibitem[Kaiser and Hilgetag(2006)]{kaiser2006}
Marcus Kaiser and Claus~C Hilgetag.
\newblock Nonoptimal component placement, but short processing paths, due to
  long-distance projections in neural systems.
\newblock \emph{PLoS Comput. Biol.}, 2\penalty0 (7):\penalty0 e95, 2006.

\bibitem[Markov et~al.(2014)]{markov2014}
N.~T. Markov et~al.
\newblock {A weighted and directed interareal connectivity matrix for macaque
  cerebral cortex}.
\newblock \emph{Cereb. Cortex}, 24\penalty0 (1):\penalty0 17--36, 2014.

\bibitem[Ercsey-Ravasz et~al.(2013)Ercsey-Ravasz, Markov, Lamy, VanEssen,
  Knoblauch, Toroczkai, and Kennedy]{ercsey-ravasz2013}
M{\'{a}}ria Ercsey-Ravasz, Nikola~T. Markov, Camille Lamy, David~C. VanEssen,
  Kenneth Knoblauch, Zolt{\'{a}}n Toroczkai, and Henry Kennedy.
\newblock {A Predictive Network Model of Cerebral Cortical Connectivity Based
  on a Distance Rule}.
\newblock \emph{Neuron}, 80\penalty0 (1):\penalty0 184--197, 2013.

\bibitem[Tropp(2015)]{tropp2015}
Joel~A. Tropp.
\newblock An introduction to matrix concentration inequalities.
\newblock \emph{Found. Trends Mach. Learn.}, 8\penalty0 (1-2):\penalty0 1--230,
  2015.
\newblock ISSN 1935-8237.

\bibitem[Cape et~al.(2017)Cape, Tang, and Priebe]{cape2017}
Joshua Cape, Minh Tang, and Carey~E. Priebe.
\newblock {The kato-temple inequality and eigenvalue concentration with
  applications to graph inference}.
\newblock \emph{Electron. J. Stat.}, 11\penalty0 (2):\penalty0 3954--3978,
  2017.
\newblock ISSN 19357524.

\bibitem[Qiu and Wicks(2014)]{qiu2014}
Robert Qiu and Michael Wicks.
\newblock \emph{{Cognitive networked sensing and big data}}, volume
  9781461445449.
\newblock 2014.
\newblock ISBN 9781461445449.

\bibitem[Oliveira(2009)]{oliveira2009}
Roberto~Imbuzeiro Oliveira.
\newblock {Concentration of the adjacency matrix and of the Laplacian in random
  graphs with independent edges}.
\newblock \emph{arXiv:0911.0600}, pages 1--46, nov 2009.

\bibitem[Preciado and Rahimian(2017)]{preciado2017}
Victor~M Preciado and M~Amin Rahimian.
\newblock Moment-based spectral analysis of random graphs with given expected
  degrees.
\newblock \emph{IEEE Transactions on Network Science and Engineering},
  4\penalty0 (4):\penalty0 215--228, 2017.

\bibitem[Le et~al.(2015)Le, Levina, and Vershynin]{le2015}
Cm~Le, E~Levina, and R~Vershynin.
\newblock {Sparse random graphs: regularization and concentration of the
  Laplacian}.
\newblock \emph{arXiv:1502.03049}, page~31, 2015.

\bibitem[Le et~al.(2017)Le, Levina, and Vershynin]{le2017}
Can~M. Le, Elizaveta Levina, and Roman Vershynin.
\newblock {Concentration and regularization of random graphs}.
\newblock \emph{Random Struct. \& Algor.}, 51\penalty0 (3):\penalty0 538--561,
  2017.
\newblock ISSN 10982418.

\end{thebibliography}
\end{document}